\documentclass[aps,prl,twocolumn,floats,showpacs,superscriptaddress]{revtex4}
\usepackage{graphicx,epsfig}
\usepackage{times}
\usepackage{graphics,dcolumn,bm,float}
\usepackage{amssymb,amsmath,rotate,color}

\begin{document}
\unitlength 1 cm
\newcommand{\be}{\begin{equation}}
\newcommand{\ee}{\end{equation}}
\newcommand{\bearr}{\begin{eqnarray}}
\newcommand{\eearr}{\end{eqnarray}}
\newcommand{\nn}{\nonumber}
\newcommand{\vk}{\vec k}
\newcommand{\vp}{\vec p}
\newcommand{\vq}{\vec q}
\newcommand{\vkp}{\vec {k'}}
\newcommand{\vpp}{\vec {p'}}
\newcommand{\vqp}{\vec {q'}}
\newcommand{\bk}{{\bf k}}
\newcommand{\bp}{{\bf p}}
\newcommand{\bq}{{\bf q}}
\newcommand{\br}{{\bf r}}
\newcommand{\bR}{{\bf R}}
\newcommand{\up}{\uparrow}
\newcommand{\down}{\downarrow}
\newcommand{\fns}{\footnotesize}
\newcommand{\ns}{\normalsize}
\newcommand{\cdag}{c^{\dagger}}

\title{Classical Analogue of the Ionic Hubbard Model}


\author{M. Hafez}
\affiliation{Department of Physics, Tarbiat Modares University, Tehran, Iran}

\author{S. A. Jafari{\footnote {Electronic address:
akbar.jafari@gmail.com}}}
\affiliation{Department of Physics, Isfahan University of
Technology, Isfahan 84156-83111, Iran}
\affiliation{School of Physics, Institute for Research in Fundamental Sciences (IPM), Tehran 19395-5531, Iran}

\author{Sh. Adibi}
\affiliation{Department of Physics, Isfahan University of
Technology, Isfahan 84156-83111, Iran}

\author{F. Shahbazi}
\affiliation{Department of Physics, Isfahan University of
Technology, Isfahan 84156-83111, Iran}


\begin{abstract}
In our earlier work [M. Hafez, {\em et al.}, Phys. Lett. A {\bf 373} (2009) 4479]
we employed the flow equation method to  obtain a classical effective model 
from a quantum mechanical parent Hamiltonian called, the ionic Hubbard model (IHM).
The classical ionic Hubbard model (CIHM) obtained in this way contains solely
Fermionic occupation numbers of two species corresponding to particles with 
$\up$ and $\down$ spin, respectively. In this paper, we employ the
transfer matrix method to analytically solve the CIHM at finite temperature 
in one dimension. In the limit of zero temperature, 
we find two insulating phases at large and small Coulomb 
interaction strength, $U$, mediated with a gap-less metallic phase, resulting in
two continuous metal-insulator transitions. Our results are further supported
with Monte Carlo simulations.
\end{abstract}
\maketitle

\section{Introduction}
To understand the magnetism and magnetic phenomena, one of the basic interactions
is the exchange mechanism which is deeply rooted in Coulomb interactions and quantum
mechanical indistinguishability. 
Therefore a fair understanding of the magnetic behavior 
of materials is not possible without investigating appropriate quantum spin models.
Introducing uni-axial anisotropy to the Heisenberg model amounts to
suppression of transverse quantum fluctuations ($S^+S^-+S^-S^+$), leading 
to the so  called Ising model~\cite{Newell}. 
The resulting Ising Hamiltonian turns 
out to contain the basic magnetic phases of the original Heisenberg
model, namely ferromagnetism, and antiferromagnetism~\cite{Feynman}. Although many 
of the interesting possible aspects such as spin liquid phases,
spin-wave excitations, etc.~\cite{Fradkin} can not
be captured by the Ising model. Classical Ising model has the 
merit of being much simpler to solve, and admits 
analytical~\cite{Newell,Onsager} and graphical solutions
~\cite{Feynman} in various geometries~\cite{JafariIJMP}
in one and two dimensions which is lacking in the original
quantum Heisenberg model.

  Dielectric properties are among the
most important properties characterizing materials. The question can
be asked here, is there any Ising-like model that can provide 
basic informations about their phase diagram, and at the same
time being simple enough to allow for analytical solutions?
We have taken the example of the 
(IHM)~\cite{Nagaosa,Egami}. This model was introduced to study the
neutral-to-ionic transition in organic compounds, as well as, 
understanding the role of strong electronic correlations in the 
dielectric properties of metal oxides~\cite{Egami,HafezCUT}.
This model is as follows:
\bearr
  H &=&-t\sum_{i\sigma}(c^{\dag}_{i\sigma}c_{i+1,\sigma}+h.c.)+
  U\sum_{i}c^{\dag}_{i\uparrow}c^{\dag}_{i\downarrow}c_{i\downarrow}c_{i\uparrow}\nn\\
  &&+\frac{\Delta}{2}\sum_{i\sigma}(-1)^{i}c^{\dag}_{i\sigma}c_{i\sigma},
  \label{ihm.eqn}
\eearr
where $c_{i\sigma}$ ($c^{\dag}_{i\sigma}$) is the usual annihilation (creation) operator
at site $i$ with spin $\sigma$. $U$ is the on-site Coulomb interaction parameter,
and $\Delta$ denotes a one-body staggered ionic potential. 
The kinetic energy scale is given by the real hopping amplitude $t$ which
prefers to gain kinetic energy by spreading the wave-function over the
whole system, leading to quantum fluctuation of the 
charge density. The zero temperature phase diagram of this model
contains Mott and band insulating states when the energy scales
corresponding to $U$ or $\Delta$ dominate, respectively~\cite{HafezCUT}.
When these two scales become of the same order of magnitude, the nature of
intermediate phase still
remains controversial. Some authors argue that the intermediate 
state is a spontaneously broken symmetry phase~\cite{SBSP},
while some others argue that the phase in between is 
metallic~\cite{Nagaosa,Bouadim2007,HafezCUT,Metals}. In our previous
investigation~\cite{HafezCUT} we employed the method of flow equations
for the quantum Hamiltonian~(\ref{ihm.eqn}) to obtain an effective
Hamiltonian in which the hopping $t$ term is renormalized to zero,
producing a new nearest neighbor Coulomb attraction.
The resulting classical Hamiltonian is of the following
lattice gas form:
\bearr
  \tilde H  =  \frac{\widetilde{\Delta}}{2}
  \sum_{i\sigma}(-1)^{i}n_{i\sigma}+
  \widetilde{U}\sum_{i}n_{i\uparrow}n_{i\downarrow} 
   +  \widetilde{V}\sum_{i\sigma,\sigma'}n_{i\sigma}n_{i\sigma'},
  \label{energy}
\eearr
where the renormalized parameters $\tilde \Delta, \tilde U,\tilde V$
have closed form expressions in terms of the bare parameters $U,\Delta$~\cite{HafezCUT}.
We take the kinetic energy scale $t$ as the unit of energy.

   By calculating the spin and charge gaps,
we showed that this simple classical model is capable of
capturing the physics of a metallic state sandwiched between two distinct
insulating phases at zero temperature as one increases $U$ for a fixed value of
$\Delta$~\cite{HafezCUT}. Here $n_{i\sigma}$ contains
two species (or colors) corresponding to $\sigma=\up,\down$, respectively.
Since $n_{\sigma}$ for each "color" $\sigma$ is either $0$ or $1$,
it is an Ising-like variable. Therefore our effective CIHM Hamiltonian
can be thought of, as two inter-penetrating Ising models on a lattice.
Being a natural extension of two species lattice gas model, 
allows for analytical solution in one spatial dimensions (1D).
Here we can employ the transfer matrix
method, constructed in terms of $3\times 3$ matrices to calculate the
thermodynamic properties of this model at finite temperatures.
At high temperatures where the thermal fluctuations wash out
the quantum effects, we expect the results of our calculations
to be a good description of the class of materials modeled in 
terms of IHM. Also a novel graphical solution in 2D similar to
Feynman's construction can be worked out~\cite{JafariUnpublished}.

The paper is organized as follows: In the next section we calculate the
grand canonical potential for this model in one dimension and discuss
the particle density and ionicity in various fillings. It is followed by a
section focused on half-filling situation, and calculate the specific heat, 
compressibility and ionicity to assess the 
nature of the phases in the parameter space of $U, T$. Throughout the 
paper we have fixed the value of $\Delta=20$~\cite{HafezCUT}.
The final section is devoted to summary and discussion.

\section{Grand Potential}
In this section we calculate the grand canonical potential that can be derived
from the grand partition function (GPF) which is defined as follows:
\be
  Z(T,\mu ,L)=\sum_{\lbrace n_{i\sigma}\rbrace}
  e^{-\beta \big( E(\lbrace n_{i\sigma}\rbrace)-\mu N(\lbrace n_{i\sigma}\rbrace)\big)},
  \label{GPF}
\ee
where $T$ is Temperature, $L$ is lattice size, and $\mu$ is chemical potential. 
The $\lbrace n_{i\sigma} \rbrace$ denotes all possible configurations of 
the occupation numbers, which must be summed over.
$N(\lbrace n_{i\sigma} \rbrace)=\sum_{i\sigma}n_{i\sigma}$
is the number of particle and $E(\lbrace n_{i\sigma} \rbrace)$ is the energy 
that is defined by the model Hamiltonian Eq.~(\ref{energy}).

  The summation in Eq.~(\ref{GPF}) can be calculated analytically as 
follows. The values of $n_{i\sigma}$ are only zero and one, which is
the Fermionic memory left in the commuting variables $n_{i\sigma}$. Hence 
$n_{i\sigma}^{2}=n_{i\sigma}$, and the second term in Eq.~(\ref{energy}) 
becomes:
\be
  \sum_{i}n_{i\uparrow}n_{i\downarrow}=
  \frac{1}{2}\sum_{i}(n_{i\uparrow}+
  n_{i\downarrow})[(n_{i\uparrow}+n_{i\downarrow})-1].
\ee
The GPF becomes,
{\setlength\arraycolsep{1pt}
\bearr
  && Z(T,\mu ,  L  )=\sum_{\lbrace n_{i\sigma}\rbrace}
  \exp \bigg\{ -\beta \Big[ \sum_{i\sigma} \big( (-1)^{i}\frac{\widetilde{\Delta}}{2}-\mu \big) n_{i\sigma} \nonumber \\
  & + & \frac{\widetilde{U}}{2}\sum_{i\sigma,\beta}n_{i\sigma}(n_{i\beta}-\frac{1}{2})
  +\widetilde{V}\sum_{i\sigma,\beta}n_{i\sigma}n_{i\beta} \Big] \bigg\}.
  \label{GPF1}
\eearr}
As can be seen at the Hamiltonian level, only the spin-summed occupations appear 
and therefore this classical model does not contain spin-polarized (magnetic) 
solutions.
So we change the summation variable from $n_{i\uparrow}$ and $n_{i\downarrow}$ 
to $n_{i}=n_{i\uparrow}+n_{i\downarrow}$ that has three possible values 0,1,2.
Therefore we have:
{\setlength\arraycolsep{1pt}
\bearr
  Z=\sum_{ \lbrace n_{i} \rbrace }\exp \bigg\{ & - & \beta \sum_{i}
  \Big[ \big( (-1)^{i}\frac{\widetilde{\Delta}}{2}-\mu \big) n_{i}
  +\frac{\widetilde{U}}{2}n_{i}(n_{i}-1) \nonumber \\
  & + & \widetilde{V}n_{i}n_{i+1}\Big] \bigg\} \prod_{i=1}^{L} (1+\delta_{n_{i},1}),
  \label{GPF2}
\eearr
}
where the coefficient $\prod_{i=1}^{L} (1+\delta_{n_{i},1})$ takes into account the
two-fold degeneracy for $n_{i}=1$ in 
Eq.~(\ref{GPF1}), which corresponds to either $n_{i\uparrow}=1$, 
$n_{i\downarrow}=0$, or $n_{i\uparrow}=0$, 
$n_{i\downarrow}=1$. Eq.~(\ref{GPF2}) can be  
written in a more symmetric form,
{\setlength\arraycolsep{1pt}
\bearr
  Z & = & \sum_{ \{ n_{i} \} } \prod_{i=1}^{L} (1+\delta_{n_{i},1})^{\frac{1}{2}}
  (1+\delta_{n_{i+1},1})^{\frac{1}{2}} \exp \bigg\{ -\beta \Big[ \nonumber \\ & + & (-1)^{i}
  \frac{\widetilde{\Delta}}{4}(n_{i}-n_{i+1})-\frac{\mu}{2}(n_{i}+n_{i+1}) \nonumber \\
  & + & \frac{\widetilde{U}}{4} \big( n_{i}(n_{i}-1)+n_{i+1}(n_{i+1}-1) \big)+
  \widetilde{V}n_{i}n_{i+1} \Big] \bigg\},
  \label{GPF3}
\eearr}
\noindent where $L$ is assumed to be even and 
the periodic boundary conditions, $n_{L+1}=n_{1}$ is implied. 
Defining the matrix elements of the transfer matrix $M_{n_{1},n_{2}}$ by,
{\setlength\arraycolsep{1pt}
\bearr
  M_{n_{1},n_{2}} & \equiv & (1+\delta_{n_{1},1})^{\frac{1}{2}}
  (1+\delta_{n_{2},1})^{\frac{1}{2}} \exp \bigg\{ -\beta \Big[ \nonumber \\ & + &
  \frac{\widetilde{\Delta}}{4}(n_{1}-n_{2})-\frac{\mu}{2}(n_{1}+n_{2}) \nonumber \\
  & + & \frac{\widetilde{U}}{4} \big( n_{1}(n_{1}-1)+n_{2}(n_{2}-1) \big)+
  \widetilde{V}n_{1}n_{2} \Big] \bigg\},
\eearr}\noindent
Eq.~(\ref{GPF3}) can be written as:
\bearr
  Z&=&\sum_{n_{1}=0}^{2} \cdots \sum_{n_{L}=0}^{2} M_{n_{2},n_{1}}M_{n_{2},n_{3}}
  \cdots M_{n_{L},n_{L-1}}M_{n_{L},n_{1}}\nonumber \\
   & = & \sum_{n_{1}=0}^{2} \cdots \sum_{n_{L}=0}^{2} M_{n_{1},n_{2}}^{t}M_{n_{2},n_{3}}
  \cdots M_{n_{L-1},n_{L}}^{t}M_{n_{L},n_{1}} \nonumber \\
  &=& \mbox{Tr}(M^{t}M)^{\frac{L}{2}},
\eearr
where $M^{t}$ is the transpose of $M$, $M_{n_{2},n_{1}}\equiv (M^{t})_{n_{1},n_{2}}$. 
Then the partition becomes:
\be
  Z=
  \lambda_{1}^{\frac{L}{2}}+\lambda_{2}^{\frac{L}{2}}+
  \lambda_{3}^{\frac{L}{2}},
\ee
where $\lambda_{1}$, $\lambda_{2}$, and $\lambda_{3}$ are the eigenvalues of 
$M^{t}M$. In the thermodynamic limit where $L$ is very large, 
the grand potential per site becomes:
\be
 \phi=-\frac{T}{2}\ln \lambda_{max}.
  \label{GP}
\ee
Here $\lambda_{max}$ is the maximum eigenvalue. Hence to obtain the grand potential 
one needs to calculate the eigenvalues of $M^{t}M$. Later on, when we discuss the 
ionicity, we will in addition need the eigenvectors  of $M^{t}M$. 
The eigenvalues of $M^{t}M$ are solutions to the third order equation,
\be
  \lambda^{3}+a_{2}\lambda^{2}+a_{1}\lambda +a_{0}=0,
  \label{3-order-E}
\ee
with $a_{2}$, $a_{1}$, and $a_{0}$ given in the appendix. 
Through the paper we will report the numerical plots for $\Delta=20$
in units in which $t=1$.
For this value of $\Delta$, there would be no level-crossing
among the eigenvalues $\lambda$ of the matrix $M^t M$ when one varies 
$\mu$, $T$, and $U$, as expected from Perron's theorem~\cite{Goldenfeld}. 
\begin{figure}[t]
  \begin{center}
    \includegraphics[angle=-90,width=8cm]{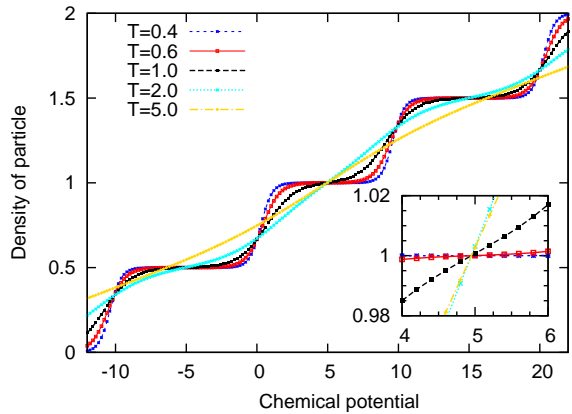}
    \caption{Particle density $n$, versus chemical potential $\mu$ for $U=10$, $\Delta=20$ 
and different temperatures. Three plateaus at $n=0.5,1,1.5$ correspond to quarter-,half-
and three quarter-fillings. For lower temperature, the variations of the density as 
a function of $\mu$ is very slow over the plateaus, which is clear signature of
a gap in the spectrum at these commensurate fillings. 
By increasing the temperature, plateaus get rounder, which
indicate the thermal energy starts to overcome the gaps.
The inset magnifies the "isosbestic" behavior at $n=1$ plateau.
}
    \label{DOP1}
  \end{center}
\end{figure}

  From the partition function, one can in principle calculate
various averages of the form $\langle n_{i\sigma}n_{j\sigma'}...\rangle$.
The simplest of these averages are the average particle density 
$n\equiv\langle n_{2i+1}+n_{2i}\rangle/2$ (symmetric combination),
and the ionicity $I\equiv\langle n_{2i+1}-n_{2i}\rangle/2$ (antisymmetric combination) 
which contain important information about the nature of the thermodynamic phases of the model.
In the following we calculate these averages as a function of $\mu$.

\subsection{Particle Density}
\label{pd.susec}
Once the grand potential $\phi$ is known, one can derive
many other thermodynamic quantities. 
The particle density, $n$, can be calculated as,
\be
  n=-\frac{\partial \phi}{\partial \mu} \Big|_{T}.
\ee

  In Fig.~\ref{DOP1}, the particle density 
versus $\mu$ has been plotted for different values of $T$. 
In this figure the value of $U$ is fixed to be $U=10$.
This value of $U$ corresponds to a band insulating regime
at zero temperature~\cite{HafezCUT}. As can be seen in the figure, 
apart from trivial cases of the empty ($n=0$) and the 
fully-filled ($n=2$) lattice, there are
three plateaus. The first one corresponds to the half-filling $n=1$ 
(i.e. one particle per lattice site), 
and two others correspond to commensurate fillings, $n=0.5$ (quarter-filling),
$n=1.5$ (three-quarter-filling) which is quite similar
to the plateaus of the parent ionic Hubbard model~\cite{Bouadim2007}.
Indeed the $n=0.5$ and $n=1.5$ are related by a particle-hole transformation.
Therefore in what follows, we focus on $n=0.5, 1$.
With increasing the temperature, the plateaus get more
and more rounded. The value of chemical potential corresponding
to the half-filling is independent of temperature, 
as can be seen in Fig.~\ref{DOP1}. The inset plot indicates
this point more transparently. This "isosbestic" behavior is observed 
for all  values of values of $10 < U < 40$.  
\begin{figure}[t]
  \begin{center}
    \includegraphics[angle=-90,width=8cm]{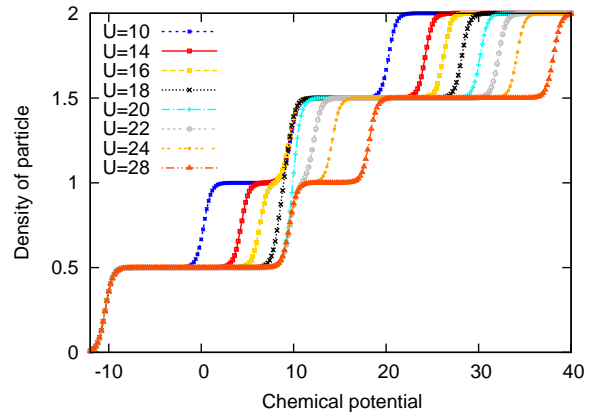}
    \caption{Particle density $n$, versus chemical potential $\mu$ 
    for constant $T=0.4,~\Delta=20$ and different values of $U$. 
    As can be seen in general again there are three plateaus corresponding to 
    $n=0.5,1,1.5$. However, for intermediate values of $U$, the plateau at 
    half-filling has disappeared.
    }
    \label{DOP2}
  \end{center}
\end{figure}

In Fig.~\ref{DOP2}, we have plotted the particle density as a function
of chemical potential $\mu$  at a constant (low) temperature $T=0.4$, 
for different values of $U$. As can be seen in the figure, increasing
$U$, causes the $n=0.5, 1.5$ plateaus which are particle-hole counterpart of
each other, get wider. However the plateau
at $n=1.0$ gets narrower and finally vanishes 
around $U=20$, resulting in a gap-less phase at half-filling.
Upon further increasing $U$, the half-filling plateau is recovered,
and gets wider, indicating the emergence of a growing new 
gap in the system, which is reminiscent of the Mott insulating behavior.
To identify the nature of gap at $n=0.5, 1.5$ one needs to
calculate the ionicity, which provides information about 
how the unit cell is being filled. This will be done in the
sequel.

  For later reference, we report the value of chemical potential
corresponding to half-filling, which by examining the numerical plots 
turns out to be:
\be
  \mu (T,U,n=1)=0.495 U.
  \label{half-filling}
\ee  
The temperature range at which the above relation is valid, is roughly $0.1\lesssim T$.
For temperatures outside this range, there will be deviations from the
above rule of thumb relation.

\subsection{Ionicity}
\label{ionicity.subsec}

\begin{figure}[t]
  \begin{center}
    \includegraphics[angle=-90,width=8cm]{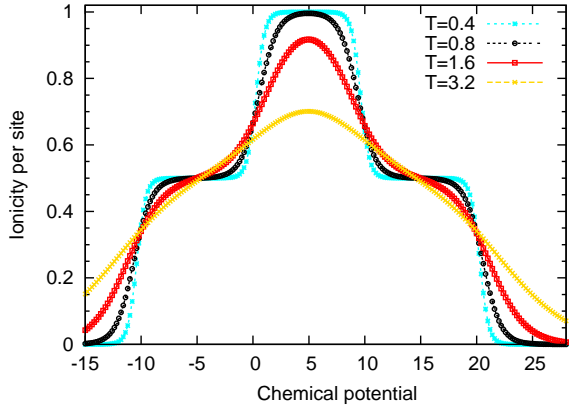}
    \caption{
    The ionicity per site as a function of chemical potential for various
    temperatures at $U=10$. Three plateaus around $\mu\approx -5,5,15$ 
    correspond to 1/4,1/2,3/4 filling.
    }
    \label{ION1}
  \end{center}
\end{figure}

If $A,B$ denote the sublattice of odd and even sites, respectively,
then $N_{A}$ and $N_{B}$ are defined as:
\bearr
  N_{A} & = & \sum_{i \in \mbox{odd}}\langle n_{i} \rangle, \\
  N_{B} & = & \sum_{i \in \mbox{even}}\langle n_{i} \rangle,
\eearr
the ionicity per site becomes $I=(N_{A}-N_{B})/L$, where $n_{i}=n_{i\uparrow}
+n_{i\downarrow}$, and $L$ is the total number of lattice sites. 
This quantity can also be calculated analytically with the aid of transfer matrices. 
In the grand canonical ensemble  $\langle n_{j} \rangle$ is given by:
\be
  \langle n_{j} \rangle =\frac{1}{Z}\sum_{\{ n_{i\sigma} \}}
  n_{j} e^{ -\beta \big( E( \{ n_{i\sigma} \})-\mu N( \{ n_{i\sigma} \}) \big)}.
\ee
Depending on whether $j$ is odd or even, 
similar to calculations for grand partition function, we obtain:
{\setlength\arraycolsep{1pt}
\bearr
  \langle n_{j} \rangle 
  & = & \frac{1}{Z}\mbox{Tr}\left(N(M^{t}M)^{\frac{L}{2}}\right),~~~~~~~~\mbox{for odd }j\\
  & = & \frac{1}{Z}\mbox{Tr}\left(N(MM^{t})^{\frac{L}{2}}\right),~~~~~~~~\mbox{for even }j
  \label{Evalue-odd}
\eearr}
where $N$ is the following $3\times3$ matrix:
\be
  N_{n_{1},n_{2}}=n_{1}\delta_{n_{1},n_{2}}.
\ee
\begin{figure}[t]
  \begin{center}
    \includegraphics[angle=-90,width=8cm]{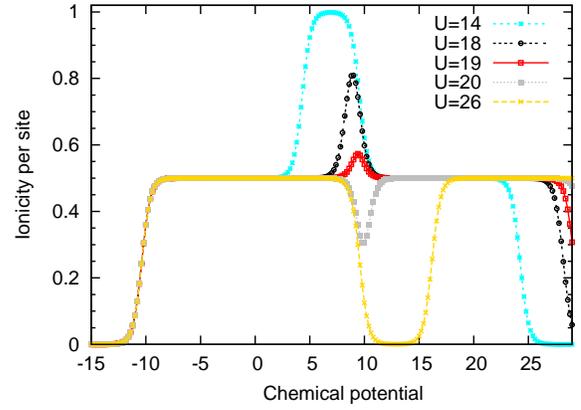}
    \caption{
    Ionicity per site as a function of $\mu$ and constant temperature $T=0.4$,
    for various values of $U$. Three plateaus correspond to 1/4,1/2,3/4 filling,
    respectively.
    Note that the center of the half-filling plateau is at $\mu=U/2$.
    }
    \label{ION2}
  \end{center}
\end{figure}

  In Fig.~\ref{ION1}, we have plotted the ionicity per site as a function
of $\mu$ for various temperatures at a fixed value of $U=10$. For this value
of $U$, at zero temperature one would expect a band insulator at half-filling,
where A-sublattice (with on-site energy $-\Delta/2$) are doubly occupied.
Let us focus around $\mu=-5$ which corresponds to quarter
filling (c.f. Fig.~\ref{DOP1}). The value of $I=0.5$ for $\mu\approx -5$
in Fig.~\ref{ION1} shows that the only electron of each unit cell
belongs to A-sublattice, which represents a charge density wave insulator.

  As we increase $\mu$ from $\approx -5$, the ionicity increases,
which indicates that sublattice A continues to be filled.
When $\mu \approx 5$, one approaches the half-filling (Fig.~\ref{DOP1}),
where at lower temperatures the A-sublattice is totally occupied, hence
$I\approx 1$. As can be seen in Fig.~\ref{ION1}, 
this saturation value is decreased as the temperature is increased. 
This reduction in the ionicity, indicates that the particles are bing thermally 
excited across  the band gap.
Further increasing of the chemical potential, one reaches the plateau
around $\mu\approx 15$ of Fig.~\ref{ION1}
(c.f. Fig.~\ref{DOP1}). The decrease in the ionicity indicates that
the added particles essentially start to occupy the B-sublattice.
The symmetry of Fig.~\ref{ION1} around $n=1$, is due to the apparent particle-hole
symmetry of the Hamiltonian.

In Fig.~\ref{ION2} we plot the ionicity per site at the constant
temperature $T=0.4$ and different values of $U$ indicated in the legend.
Again there are three plateaus corresponding to 1/4,1/2 and 3/4-filling,
respectively. This figure indicates that the center of the half-filling
plateau is at $\mu_{1/2}=U/2$. This can be understood as a Hartree like
energy for the Hubbard model. 
 Now concentrating around half-filling in this figure, for small values of $U$, the ionicity 
reaches $1$, which indicates a complete electric polarization of the unit cell, and hence 
we have a band insulator. For intermediate values of $U$, the maximum value
of ionicity does not reach $1$, which shows some of the added particles 
start to occupy the B-sublattice. Also note that for the intermediate values
of $U$ the  half-filling plateau starts to disappear, i.e. the emergence
of a metallic behavior. 
For large values of $U$, the half-filling plateau is restored, but at
zero ionicity. Therefore we have an insulator with unpolarized unit cell.
Such an insulating state can be thought of as classical counterpart of
the Mott insulating state.

\section{Half-filling}
\begin{figure}[t]
  \begin{center}
    \includegraphics[angle=-90,width=8cm]{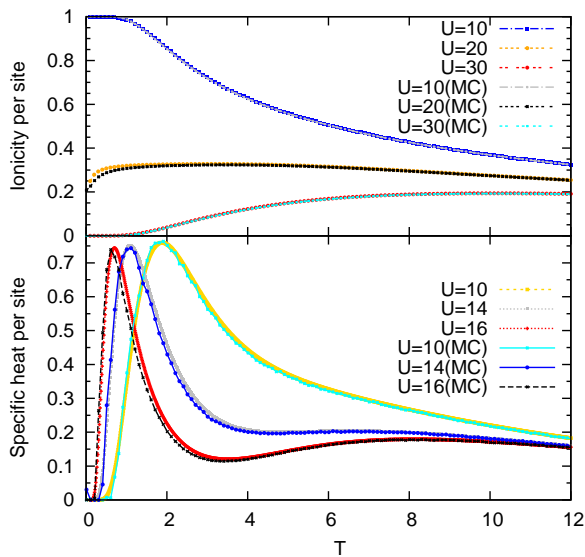}
    \caption{Ionicity and specific heat per site at half-filling 
    from transfer matrix and Monte Carlo (MC) methods. To stay in the
    range of validity of Eq.~(\ref{half-filling}), transfer matrix results 
    are plotted for $T \geq 0.1$. Monte Carlo results are obtained for 
    $L=500$ sites. As stated earlier, we are working at constant $\Delta=20$.
    } 
    \label{compare}
  \end{center}
\end{figure}

As we saw in Fig.~\ref{DOP2}, for $n=0.5, 1.5$ 
the  $U,\Delta$ energy scales cooperate with each other, to give rise
to a charge density wave insulating ground state.
Rather at $n=1.0$, these two energy scales compete against each other
to destroy the insulating behavior for intermediate $U$ ($\sim\Delta$), giving
rise to a richer phase diagram. Therefore in this section we focus at
half-filling and calculate the specific heat, compressibility, 
and ionicity. Before doing so, we compare some physical
quantities evaluated by a fully numerical Monte Carlo simulation,
with our exact transfer matrix results. In Fig.~\ref{compare}, ionicity 
and specific heat per site are plotted at half-filling and show a good 
agreement between analytical and numerical results. This ensures that 
both Monte Carlo and analytic results are reliable.

Now let us proceed with the calculation of various thermodynamic quantities.
For the fixed value of $\Delta=20$, we have two ways to plot thermodynamic
quantities. First way is to plot them as a function of temperature $T$, at
some selected values of $U$. These results indicate that in the present one
dimensional model, there will be no finite temperature phase transition. 
This is obviously due to the analytic behavior of the partition function
as a function of $T$.
The second
way is to plot them as a function of $U$, for some selected temperatures. 
This second way of presenting the data, reveals that as one lowers the temperature,
there will be sharper features as a function of $U$, indicating the zero 
temperature  phase transition.
Our calculations are based on Eqs.~(\ref{GP},\ref{half-filling}). 
For very low temperatures, where the validity of Eq.~(\ref{half-filling}), 
might be questioned, we employ Monte Carlo simulation data.

\subsection{Specific heat}
\begin{figure}[t]
  \begin{center}
    \includegraphics[angle=-90,width=8cm]{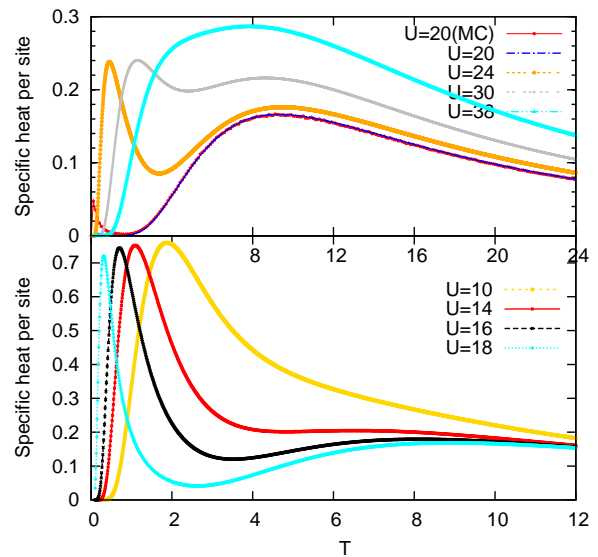}
    \caption{Specific heat per site versus $T$ at $\Delta=20$ for different values of $U$. 
    For small values of $U$ there is a single peak, which becomes a peak-dip-hump
    in intermediate $U$, and finally the peak merges into the hump at large $U$.
    For $U=20$, both analytic and Monte Carlo results are reported .
    MC results are for $L=500$ sites. 
    }
    \label{SHT1}
  \end{center}
\end{figure}
Specific heat per site can be calculated as:
\be
  c_{L}=T\frac{\partial s}{\partial T},
\ee
where $s$ is entropy per site at half-filling that can be derived from
the grand potential.
Figure~\ref{SHT1} shows the specific heat versus $T$ for various
values of $U$. As can be seen in Fig.~\ref{SHT1} for values
of $U\le 14$, there is a single peak in the $c_L$. 
For $14<U<38$ a peak-dip-hump structure can be observed. 
For $U>20$ the hump is quite clear, while for $U<20$,
it can be interpreted as a precursor to the hump.
For $38\le U$, the peak merges into the  hump structure.
In terms of the parent quantum Hamiltonian, such hump structure
indicates incoherent excitations.
For $U=20$, Eq.~(\ref{half-filling})
is not reliable at very low temperatures. Therefore we report Monte
Carlo simulation results which indicates a very sharp peak at $T\approx 0$.
As $U$ moves towards $\approx 20$ from both sides (lower and upper
panels), the peak gets sharper and moves towards lower temperatures.
This indicates that in $T\to 0$ limit one expects a transition
between gaped and gap-less states. 
\begin{figure}[t]
 \begin{center}
    \includegraphics[angle=-90,width=8cm]{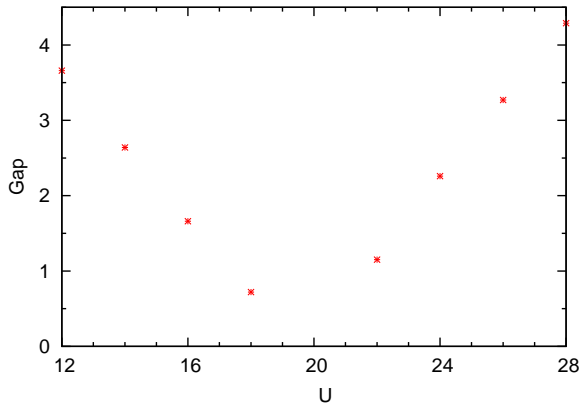}
    \caption{Estimate of gap using two-state model as a function
    of $U$ at $\Delta=20$.
    }
    \label{gap}
  \end{center}
\end{figure} 
This behavior can be simply understood in terms of a two-state 
model with level spacing $\delta$, whose specific heat is given by
\be
   c_{L}=k_B\left(x^2\frac{{\rm e}^x}{(1+{\rm e}^x)^2}\right),
   \label{cl2lev}
\ee 
where $x=\delta/(k_BT)$. Behavior of Eq.~(\ref{cl2lev}) for $x\gg 1$ is
like $\sim x^2{\rm e}^{-x}$, while for $x\ll 1$ it vanishes as $\sim x^2$. 
For the intermediate region a Schottky peak around 
$x_{\rm peak}\sim 1$ ($\delta\sim k_BT$) arises in the specific heat.
Fitting the specific data to Eq.~(\ref{cl2lev}), in Fig.~\ref{gap} we have 
plotted the estimated gap versus $U$, which indicates two gaped phases.
According to the above two-state formula, this peak corresponds to
$\delta/k_BT\sim 1$, from which a gap of $\delta\sim 10^{-2}$ can
be estimated.
If we extrapolate the estimated gap for $18<U<22$, zero gap region is expected to occur
for $U_{c_1}=19.30 \lesssim U \lesssim U_{c_2}=19.75$ 
which is compatible with our previous work~\cite{HafezCUT}. 
\begin{figure}[b]
 \begin{center}
    \includegraphics[angle=-90,width=8cm]{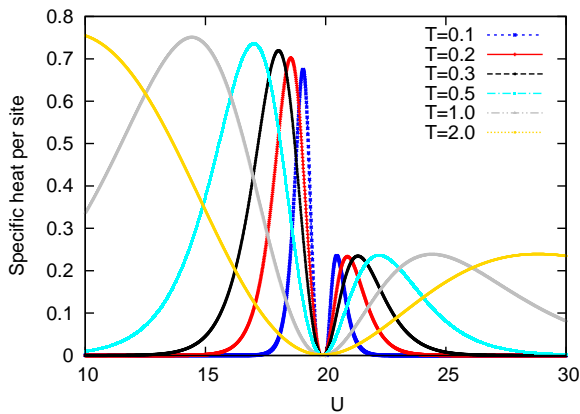}
    \caption{Specific heat per site versus $U$ for different values 
    of $T$.}
    \label{SHT2}
  \end{center}
\end{figure}  

   Now let us look at the specific heat data from a different angle.
In Fig.~\ref{SHT2} we plot the specific heat versus $U$ for different
values of temperatures. As can be seen, there are two peak structures
at all temperatures, which get sharper and by lowering the temperature,
they tend to accumulate around $U\approx 20$. 
Extrapolating the trend of this double-peak 
structure to the limit of $T\to 0$, suggests two phase transitions at 
$U_{c1}$ and $U_{c2}$~\cite{HafezCUT}, compatible with the behavior 
of the vanishing gap region in Fig.~\ref{gap}.
The characteristic quadratic behavior around $U\approx 20$ seen in Fig.~\ref{SHT2}, 
which according to the two-state model is expected to be like $c_L\propto (\delta/T)^2$,
indicates two continuous metal-insulator transitions, with 
$\delta \sim |U-U_{c_i}|$, $i=1,2$.

\subsection{Compressibility}
  Another quantity that can be treated in our consideration 
CIHM is the compressibility which is given by:
\be
  \kappa=
  \frac{1}{n^{2}}\frac{\partial n}{\partial \mu}\Big|_{T},
  \label{compressibility}
\ee
where $n=N/L$ is the density of particles per site.
\begin{figure}[t]
  \begin{center}
    \includegraphics[angle=-90,width=8cm]{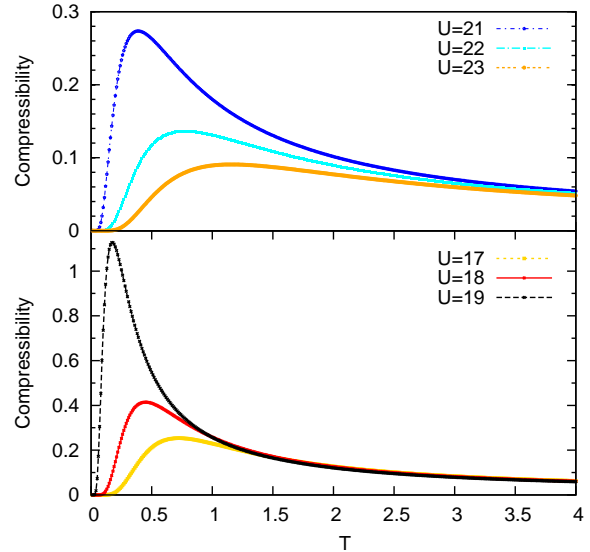}
    \caption{Compressibility versus $T$ for different values of $U$.}
    \label{CT}
  \end{center}
\end{figure}

   In Fig.~\ref{CT} we plot the compressibility as a function of $T$ for 
different values of $U$. Zero compressibility is a characteristic of 
gaped states. As can be seen in this figure for small value of $U$ the range
of temperatures at which the compressibility is close to zero is substantial,
which means that the gap is so large that up to such temperature the insulating
behavior is still manifest. By increasing $U$, this temperature range shrinks
and becomes smaller and smaller, until around $U=20$, it extrapolates to zero.
Increasing $U$ beyond $20$, again recovers a finite temperature range in which the
compressibility is zero. This behavior confirms that a gap-less state is 
sandwiched between two gaped states.

\begin{figure}[t]
  \begin{center}
    \includegraphics[angle=-90,width=8cm]{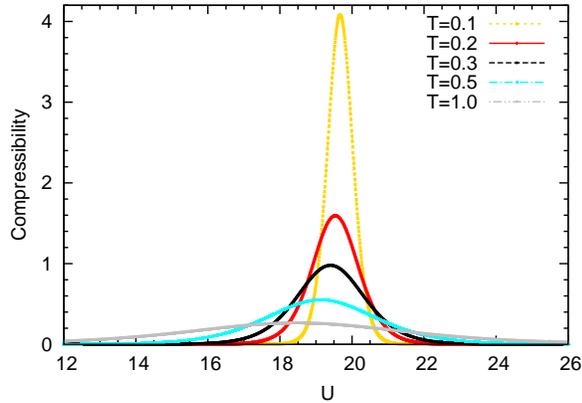}
    \caption{Compressibility versus $U$ for 
different values of temperature.}
    \label{CU}
  \end{center}
\end{figure}

To see the above statement more clearly, in Fig.~\ref{CU} we plot compressibility
as a function of $U$ for selected temperatures.
As can be seen in this figure, there is a region with non-zero compressibility,
which characterizes a gap-less phase. Outside this region, the compressibility 
decays to zero. By decreasing $T$, the width of the compressibility peak becomes
smaller and while the height of the peak diverges as $T\to 0$; a typical 
characteristic of a continuous phase transition.
This confirms the existence of a metallic phase at zero temperature~\cite{HafezCUT}.
The effect of thermal fluctuations is to smear the edges of metallic region.
This is quite intuitive, as for values of $U$ near the zero temperature 
boundary of metallic phase with neighboring insulating phases, the gaps 
are small, and hence the thermal energy can overcome the gap.

\subsection{Ionicity}

\begin{figure}[b]
  \begin{center}
    \includegraphics[angle=-90,width=8cm]{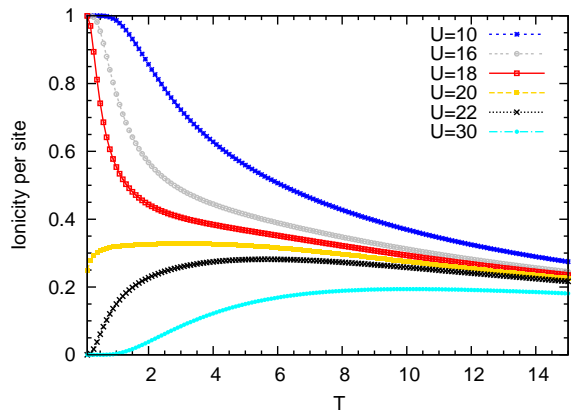}
    \caption{Ionicity versus $T$ for $\Delta=20$ and different values of $U$ 
    at half-filling.}
    \label{I-versus-T}
  \end{center}
\end{figure}

  In Fig.~\ref{I-versus-T} we plot the ionicity per site 
for $\Delta=20$ and various values of $U$ at half-filling.
As can be seen in the figure, by lowering the temperature,
for $U\le 18$, the ionicity tends to $I=1$; 
for $U>20$ it reaches a zero temperature value of $I=0$; 
while for $U\approx 20$, it reaches a value between these two limits; 
characterizing a phase with charge fluctuations.

This behavior can be understood as follows: At small $U$
regime the unit cell is fully polarized at low temperatures, with
both $\up$ and $\down$ particles occupying the A-sublattice. As the 
temperature is increased, some of the particles get excited to 
B-sublattice by absorbing the thermal energy, $k_BT$.
Similarly for large values of $U$, at lower temperatures the 
unit cell is not polarized, due to the Coulomb term $U$. By
increasing the temperature, thermal excitations with doubly
occupied sites will be created, thereby increasing the ionicity.
For intermediate $U\approx 20$, the weights of  polarized and unpolarized
configurations in the unit cell become comparable; hence giving
the ionicity $0<I<1$.

\begin{figure}[t]
  \begin{center}
    \includegraphics[angle=-90,width=8cm]{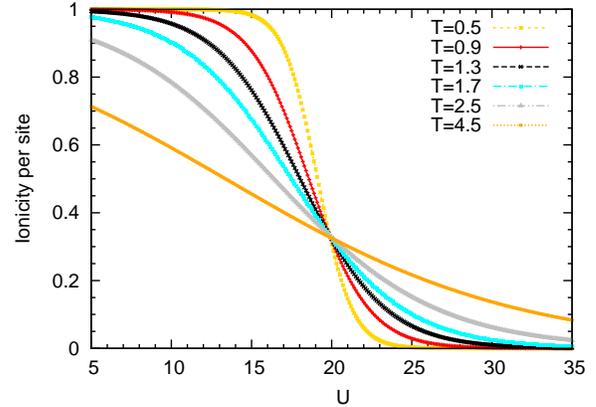}
    \caption{Ionicity versus $U$ for some values of temperature at half-filling
    and $\Delta=20$.
    }
    \label{I-versus-U}
  \end{center}
\end{figure}

In Fig.~\ref{I-versus-U} we plot the ionicity at half-filling as
a function of $U$ for $\Delta=20$ and various values of $T$.
As can be seen at all temperatures, the ionicity smoothly varies between 
$1$ for small values $U$, and $0$ for large values of $U$. 
The width of the transition region decreases by lowering the temperature,
and is expected to approach the figure 2 of Ref.~\onlinecite{HafezCUT}.

\section{Conclusion}
In this work we studied a classical model consisting of two Ising like
variables on a one dimensional chain. Despite the simplicity which essentially
results form the lack of Fermionic minus sign problem, our model captures
some of the interesting properties of the ionic Hubbard model.
Various thermodynamic properties, such as, specific heat, ionicity, 
particle density and compressibility when viewed as a function of $U$
in a given temperature, indicate the presence of two gaped states
at small and large values of $U$, with a gap-less state sandwiched
between them (around $U/\Delta\approx 1$). When the same quantities viewed as a function of 
temperature, there is no sign of phase transition down to zero
temperature. The three phase scenario of the zero temperature 
with clear zero temperature phase transition boundaries at $U_{c_1}(\Delta)$ 
and $U_{c_2}(\Delta)$ extrapolates to higher temperatures. However,
the boundaries get smeared due to thermal fluctuations, giving rise
to a cross-over behavior.

  Mapping of $D$ dimensional quantum models to 
$D+1$ dimensional classical Hamiltonians is a well known paradigm
in statistical physics. Our flow equation approach~\cite{HafezCUT} suggests an
alternative approach to construct "$D$" dimensional classical
models which might be useful in capturing basic aspects of the original 
quantum Hamiltonian. 

\section{Acknowledgements}
We would like to thank Professor N. Nafari for useful comments and critique of 
the manuscript.  
This research was supported by the Vice Chancellor for 
Research Affairs of the Isfahan University of Technology (IUT). 
S.A.J was supported by the National Elite Foundation (NEF) of Iran.

\section{Appendix}
\subsection{Coefficients of Eq.~(\ref{3-order-E})}
The coefficients     $a_{2}$, $a_{1}$, and $a_{0}$ of Eq.~(\ref{3-order-E})
are as follows:
{\setlength\arraycolsep{1pt}
\bearr
  a_{2} \equiv & - & \{ {{\rm e}^{-\beta\, \left( 2\, \widetilde{\Delta} +2\,\mu+ \widetilde{U}  \right) }}+2\,{
{\rm e}^{-3/2\,\beta\, \left(  \widetilde{\Delta} +2\,\mu \right) }} + {{\rm e}^{-
\beta\, \left( 4\,\mu+ \widetilde{\Delta}  \right) }} \nn \\ & + & 2\,{{\rm e}^{-1/2\,\beta\,
 \left( 3\, \widetilde{\Delta} +2\,\mu+2\, \widetilde{U} +8\, \widetilde{V}  \right) }}+4\,{{\rm e}^{-\beta\,
 \left(  \widetilde{\Delta} +2\,\mu+2\, \widetilde{V}  \right) }} \nn \\ &+& 2\,{{\rm e}^{-1/2\,\beta\,
 \left(  \widetilde{\Delta} +6\,\mu \right) }}+{{\rm e}^{-\beta\, \left( 2\, \widetilde{U} +8\, \widetilde{V} +
 \widetilde{\Delta}  \right) }}+{{\rm e}^{-\beta\, \left(  \widetilde{U} +2\,\mu \right) }} \nn \\ &+& 
2\,{{\rm e}^{-1/2\,\beta\, \left( 2\, \widetilde{U} +8\, \widetilde{V} +2\,\mu+ \widetilde{\Delta}  \right) }}
 \} {{\rm e}^{\beta\, \left( 4\,\mu+ \widetilde{\Delta}  \right) }}, 
\eearr}
{\setlength\arraycolsep{1pt}
\bearr
  a_{1} & \equiv & - \{ -4\,{{\rm e}^{-\beta\, \left( 2\, \widetilde{\Delta} +2\,\mu+ \widetilde{U} +2\, \widetilde{V} 
 \right) }}+4\,{{\rm e}^{-1/2\,\beta\, \left( 4\, \widetilde{U} +12\, \widetilde{V} +2\,\mu+ \widetilde{\Delta} 
 \right) }} \nn \\ &-& 4\,{{\rm e}^{-\beta\, \left( 2\, \widetilde{U} +10\, \widetilde{V} + \widetilde{\Delta}  \right) }}-
2\,{{\rm e}^{-1/2\,\beta\, \left( 4\, \widetilde{U} +16\, \widetilde{V} +3\, \widetilde{\Delta} +2\,\mu \right) 
}} \nn \\ &-& {{\rm e}^{-\beta\, \left( 2\, \widetilde{U} +8\, \widetilde{V} +2\,\mu+ \widetilde{\Delta}  \right)}}+4\,{
{\rm e}^{-1/2\,\beta\, \left( 2\, \widetilde{U} +4\, \widetilde{V} + \widetilde{\Delta} +6\,\mu \right) }} \nn \\ &-& {
{\rm e}^{-\beta\, \left( 2\, \widetilde{U} + \widetilde{\Delta} +2\,\mu \right) }}+4\,{{\rm e}^{-1
/2\,\beta\, \left( 4\, \widetilde{U} +12\, \widetilde{V} +3\, \widetilde{\Delta} +2\,\mu \right) }} \nn \\ &+& 4\,{{\rm e}^
{-1/2\,\beta\, \left( 2\, \widetilde{U} +3\, \widetilde{\Delta} +6\,\mu+4\, \widetilde{V}  \right) }}-2\,{
{\rm e}^{-1/2\,\beta\, \left(  \widetilde{\Delta} +6\,\mu+2\, \widetilde{U}  \right) }} \nn \\ &+& 8\,{
{\rm e}^{-\beta\, \left( 2\, \widetilde{\Delta} +2\,\mu+ \widetilde{U} +3\, \widetilde{V}  \right) }}+8\,{
{\rm e}^{-\beta\, \left( 4\,\mu+ \widetilde{\Delta} + \widetilde{V}  \right) }}-4\,{{\rm e}^{-\beta\, \left( 4\,\mu+
 \widetilde{\Delta}  \right) }} \nn \\ &-& 4\,{{\rm e}^{-
\beta\, \left( 2\, \widetilde{U} +8\, \widetilde{V} + \widetilde{\Delta}  \right) }}-2\,{{\rm e}^{-1/2\,\beta\,
 \left( 4\, \widetilde{U} +3\, \widetilde{\Delta} +8\, \widetilde{V} +2\,\mu \right) }} \nn \\ &+& 8\,{{\rm e}^{-\beta\,
 \left( 2\, \widetilde{U} +9\, \widetilde{V} + \widetilde{\Delta}  \right) }}-2\,{{\rm e}^{-1/2\,\beta\, \left( 
4\, \widetilde{U} +8\, \widetilde{V} + \widetilde{\Delta} +2\,\mu \right) }} \nn \\ &+& 2\,{{\rm e}^{-\beta\, \left( 2\, \widetilde{U} +4
\, \widetilde{V} + \widetilde{\Delta} +2\,\mu \right) }} -4\,{{\rm e}^{-\beta\, \left( 2\, \widetilde{\Delta} +2\,\mu+ \widetilde{U} +4\, \widetilde{V} 
 \right) }} \nn \\ &-& 4\,{{\rm e}^{-\beta\, \left(  \widetilde{U} +2\, \widetilde{V} +2\,\mu \right) }}-2\,{
{\rm e}^{-1/2\,\beta\, \left( 2\, \widetilde{U} +3\, \widetilde{\Delta} +6\,\mu+8\, \widetilde{V}  \right) }} \nn \\ &-& 2
\,{{\rm e}^{-1/2\,\beta\, \left( 4\, \widetilde{U} +16\, \widetilde{V} +2\,\mu+ \widetilde{\Delta}  \right) }}-2
\,{{\rm e}^{-1/2\,\beta\, \left( 2\, \widetilde{U} +6\,\mu+8\, \widetilde{V} + \widetilde{\Delta}  \right) }} \nn \\ &-& 2
\,{{\rm e}^{-1/2\,\beta\, \left( 2\, \widetilde{U} +3\, \widetilde{\Delta} +6\,\mu \right) }}-4\,{
{\rm e}^{-\beta\, \left( 4\,\mu+ \widetilde{\Delta} +2\, \widetilde{V}  \right) }} \nn \\ &+& 8\,{{\rm e}^{-
\beta\, \left(  \widetilde{U} +3\, \widetilde{V} +2\,\mu \right) }}-4\,{{\rm e}^{-\beta\, \left(  \widetilde{U} 
+4\, \widetilde{V} +2\,\mu \right) }} \} {{\rm e}^{\beta\, \left(  \widetilde{\Delta} +6\,\mu
 \right) }}, 
\eearr}
\bearr
  a_{0} & \equiv & 4\,{{\rm e}^{-2\,\beta\, \left(  \widetilde{U} + \widetilde{V} -3\,\mu \right) }} \{ -4\,{
{\rm e}^{-2\,\beta\, \widetilde{V} }}-4\,{{\rm e}^{-5\,\beta\, \widetilde{V} }}-4\,{{\rm e}^{-3\,
\beta\, \widetilde{V} }} \nn \\ &+& 10\,{{\rm e}^{-4\,\beta\, \widetilde{V} }}+4\,{{\rm e}^{-\beta\, \widetilde{V} }}+4\,{
{\rm e}^{-7\,\beta\, \widetilde{V} }}-4\,{{\rm e}^{-6\,\beta\, \widetilde{V} }} \nn \\ &-& {{\rm e}^{-8\,
\beta\, \widetilde{V} }}-1 \} . 
\eearr

\end{document}